\documentclass[12pt]{article}
\usepackage{amssymb}
\usepackage[dvips]{epsfig}
\usepackage[utf8]{inputenc}
\usepackage{graphicx}
\usepackage{amsmath} 

\def \del{\partial}

\newcommand{\be}{\begin{equation}}
\newcommand{\ee}{\end{equation}}
\def\bea{\begin{eqnarray}}
\def\eea{\end{eqnarray}}
\newcommand{\bn}{\begin{eqnarray}}
\newcommand{\en}{\end{eqnarray}}
\newcommand{\nn}{\nonumber}
\newcommand{\no}{\noindent}

\newcommand{\HD}{H_{\mu\nu}^{Diff}}
\newcommand{\HTD}{H_{\mu\nu}^{TD}}
\newcommand{\HWT}{H_{\mu\nu}^{WT}}

\newcommand{\p}{\partial}
\def\bea{\begin{eqnarray}}
\def\eea{\end{eqnarray}}
\newcommand{\beq}{\begin{eqnarray}}
\newcommand{\eeq}{\end{eqnarray}}

\def \Lag {\mathcal{L}}
\def \Lagstm {\mathcal{L}_m^{ST}}
\def \Lagsto {\mathcal{L}_0^{ST}}

\def \tr {\tilde{R}}

\begin{document}

\title{\textbf{The dimensional reduction of linearized spin-2 theories invariant under transverse diffeomorphisms}}
\author{D. Dalmazi$^{1}$\footnote{denis.dalmazi@unesp.br}  and R. R. Lino dos Santos$^{1,2}$\footnote{rado@cp3.sdu.dk} \\
\textit{{1- UNESP - Campus de Guaratinguet\'a - Departamento de F\'isica}}\\
\textit{{CEP 12516-410 - Guaratinguet\'a - SP - Brazil.} }\\
\textit{{2- CP3-Origins, University of Southern Denmark}}\\  
\textit{{Campusvej 55, DK-5230 Odense M, Denmark.}} }
\date{\today}
\maketitle

\begin{abstract}

Here we perform the Kaluza-Klein dimensional reduction from $D+1$ to $D$ dimensions of massless Lagrangians described by a symmetric rank-2 tensor and invariant under transverse differmorphisms (TDiff). They include the linearized Einstein-Hilbert theory, linearized unimodular  gravity and scalar tensor models. We obtain simple expressions in terms of gauge invariant field combinations and show that unitarity is preserved in all cases. After fixing a gauge,  the reduced model becomes a massive scalar tensor theory. We show that the diffeomorphism (Diff) symmetry, instead of TDiff, is a general feature of the massless sector of consistent massive  scalar tensor models. We discuss some subtleties when eliminating Stückelberg fields directly at action level as gauge conditions.% We also show that the reduced models all have a smooth massless limit. 

A non local connection between the massless sector of the scalar tensor theory and the  pure tensor TDiff  model leads to a parametrization of the non conserved source which naturally separates spin-0 and spin-2 contributions in the pure tensor theory.
 
The case of curved backgrounds is also investigated.
If we truncate the non minimal couplings to linear terms in the curvature, vector and scalar constraints require Einstein spaces  as in the Diff and WTDiff (Weyl plus Diff) cases. We prove that our linearized massive scalar tensor models admit those curved background extensions.

\end{abstract}

\newpage

\section{Introduction}

The Einstein general relativity (GR) and the standard cosmological model $\Lambda CDM$  are very successful despite our present lack of understanding on important issues like the cosmological constant, dark energy, dark matter,  and the discrepancy between different measurements of the Hubble constant, see \cite{lh} for a recent review work with more details.
One of the possible modifications of GR which would be relevant for  large scale physics and eventually contribute to our understanding of late time acceleration and the cosmological constant, is the existence of some graviton mass ($m_{gr}$). One of the striking successes of GR is the detection of gravitational waves coming from black-hole and neutron stars mergers \cite{abbott}, those experiments set an upper bound of about $10^{-23}$ ev  for $m_{gr}$ which is very small but it does not eliminate a possible tinier graviton mass still relevant for the accelerated expansion of the Universe in a pure massive gravity scenario, see \cite{hinter,massive} for review works on massive gravity.

Moreover, in order to figure out why the graviton is eventually  massless we must, of course, start with some non zero mass and search for its consequences.  In the early 1970$'$s,  one \cite{vdvz} noticed  that no matter how small is $m_{gr}$, its consequences for solar system tests of the gravitational interaction would be disastrous. This is the known vDVZ mass discontinuity problem. Soon one has realized \cite{vain} that the graviton mass introduces another scale in the theory that spoils the linearized approximation used in \cite{vdvz} as $m_{gr}\to 0$. Therefore, non linear terms must be considered, however they lead to ghosts \cite{ghost}. That problem remained for decades until  one has finally found  graviton potentials which eliminate the ghost and solve the vDVZ problem altogether, see \cite{drgt} and \cite{rosen}  and also the
review works \cite{hinter,massive}. A general fiducial (fixed) metric has been introduced in \cite{hs2} and later the addition of a kinetic term for such metric gave rise to the bimetric model \cite{hs3} which
has a massive plus a massless graviton coupled to each other, see also the review work \cite{bimetric}. The bimetric model does not suffer from an unstable Friedmann-Lemaitre-Robertson-Walker (FLRW) solution, a problem present in the model of \cite{drgt}, see \cite{defelice}. Moreover, in the bimetric model \cite{hs3} the graviton mass does not need to be small, see \cite{hm} for recent detailed studies on observational constraints on the parameter space of the bimetric model which includes the Fierz-Pauli mass.

In earlier calculations \cite{vdvz,vain,ghost}, in the ghost free massive gravities \cite{drgt,hs2} and also in the bimetric model
\cite{hs3}, the massive Fierz-Pauli (FP) model \cite{fp} is the paradigmatic starting point. It is a free theory for massive spin-2 particles where the linearized diffeomorphism (Diff) symmetry of the massless sector, linearized Einstein-Hilbert (LEH), is broken by the mass terms. Since the minimal symmetry for massless spin-2 particles is transverse diffeomorphisms (TDiff) instead of Diff, see \cite{van}, it is natural to search for the minimal way of describing massive spin-2 particles by adding mass terms to a massless TDiff  tensor model. The question is: Is TDiff the minimal symmetry of the massless sector of a massive spin-2 theory ? This has been investigated in \cite{blas}. The TDiff tensor model describes in general massless spin-2 and spin-0 particles. The authors of \cite{blas} have concluded that, although one could add a consistent mass term for the spin-0 sector without problems, there is no mass term in the spin-2 sector that might avoid the presence of ghosts. They have shown that such negative result holds also in the special case of the linearized unimodular gravity where  the TDiff symmetry is enhanced to WTDiff (Weyl plus TDiff). 

Instead of trying to obtain mass terms via an Ansatz one can follow another route and automatically generate them by means of a Kaluza-Klein \cite{kaluzaklein} dimensional reduction from $D+1$ to $D$ dimensions, restricted to one massive mode. Such procedure allows us  to go, for instance,  from Maxwell to Maxwell-Proca, from LEH to massive FP and holds also for higher spins \cite{rss,bhr}. This has been done in \cite{bfh} in the special case of the WTDiff model.  We comment on the results of \cite{bfh} and compare to ours whenever possible. In particular, we end up with simpler compact formulae for the reduced massive theory and clarify the issue of gauge fixing, which is  subtle now due to the transverse condition on the vector symmetry of the massless theory.

 Here we perform the KK dimensional reduction of an arbitrary TDiff tensor model in order to generate possible mass terms and look for generalizations of the resulting models in curved backgrounds, see sections 3 and 5 respectively. 
In section 4 we analyse its content and present a covariant proof of unitarity. We find an interesting parametrization of the non conserved source in the massless TDiff tensor case. In section 6 we draw our conclusions, stressing that the massless limit must have Diff symmetry instead of TDiff. If we insist in TDiff symmetry at $m=0$ we have no possible mass term. The appendix shows, for the reader's convenience, some technical details about spin projection and transition operators used to write down propagators.

\section{The model and the notation}

Here we  closely follow the notation of \cite{blas} but we use $\eta_{\mu\nu}=diag(-1,1,...,1)$.
In terms of a rank-2 symmetric tensor $h_{\mu\nu}$ we start from the second order Lagrangian in  $D \ge 3$ dimensions
\be 
\Lag(a,b)=-\frac{1}{4}\del_\mu h^{\alpha\beta}\del^\mu h_{\alpha\beta}+\frac{1}{2}\del^\mu h^{\alpha\beta} \del_{\alpha}h_{\mu\beta}-\frac{a}{2}\;\del^\mu h \del^\nu h_{\mu\nu}+\frac{b}{4}\del_\mu h \del^\mu h \quad .
\label{L2}
\ee
\no The first term is required for spin-2 propagation and the second one must be as it stands for the  elimination of  spin-1 ghosts \cite{blas}. For any $a,b\in\mathbb{R}$, $\Lag(a,b)$ is invariant under TDiff  ($\del^\mu \xi^T_\mu=0 $):
\be 
\delta h_{\mu\nu} = \del_\mu \xi^{T}_{\nu} +  \del_\nu \xi^{T}_{\mu},
\label{tdiffs}
\ee

 It is convenient to split the discussion in three cases. The TDiff symmetry can be enhanced either to WTDiff (Weyl plus TDiff) or to arbitrary diffeomorphisms (Diff) according to a fine tune of the coefficients $(a,b)$. Regarding the first case, the reader can check that the  Weyl transformation $\delta h_{\mu\nu} = \Lambda \eta_{\mu\nu}$ will be a symmetry of \eqref{L2} only if  $(a,b)=\left(2/D,(D+2)/D^2\right)$. The corresponding WTDiff model $\Lag \left(2/D,(D+2)/D^2\right)$ is the linearized version of unimodular gravity \cite{blas,bfh}. The second case corresponds to $a=1=b$. Now the symmetry is enhanced to arbitrary linearized diffeomorphisms (Diff): $\delta h_{\mu\nu} = \del_\mu \xi_\nu + \del_\nu \xi_\mu $. In this case  $\Lag(1,1)=\Lag_{LEH}$ is the linearized version of the Einstein-Hilbert theory $\kappa\sqrt{-g}R$ about flat background, $g_{\mu\nu} = \eta_{\mu\nu} + h_{\mu\nu}/\sqrt{\kappa}$. It turns out that $\Lag(1,1)$ is just one example of a one parameter family of equivalent Lagrangians. Through the invertible  redefinition ($\lambda$-shift) $h_{\mu\nu}\rightarrow  h_{\mu\nu}+\lambda\; h \eta_{\mu\nu}$, with $\lambda\neq-1/D$, we generate a family of models $\Lag(A(\lambda),B(\lambda))$ of the same form (\ref{L2}), where

\bea
A(\lambda)&=& a+\lambda(D\;a-2),\label{a}\\
B(\lambda)&=& b+2\lambda(D\;b-a-1)+\lambda^2 [D^2 b-D(2a+1)+2].
\label{b}
\eea
In particular, if we start with the linearized Einstein-Hilbert theory    $(a,b)=(1,1)$, we  generate a Diff family invariant under  $\delta h_{\mu\nu} = \del_\mu \xi_\nu + \del_\nu \xi_\mu  -2\lambda (\p^{\rho}\xi_{\rho})/(\lambda\, D +1)$. Notice that there is no  WTDIFF family since $(a,b)=\left(2/D,(D+2)/D^2\right)$ is a fixed point of the $\lambda$-shift. It turns out that for both the WTDiff model and the Diff family, the Lagrangian $\Lag(a,b)$ describes physical massless spin-2 particles. Moreover, only in those cases we have $f_D(a,b)=0$, where we have defined a function of the parameters of the TDiff model in $D$ dimensions:
\be 
f_{D}(a,b) \equiv 1-2a+a^2(D-1)-b(D-2).
\label{d0}
\ee 
The above quantity shows up after eliminating $\lambda$ from \eqref{a} and plugging it in \eqref{b}, with $(a,b)=(1,1)$; accordingly we arrive at   $f_D(A(\lambda),B(\lambda))=0$. 

The third case is the main concern here and corresponds to pure TDiff models without enhanced symmetry. They describe massless spin-2 and massless spin-0 particles, the former are always physical while the latter ones are physical (definite positive Hamiltonian) whenever, see \cite{blas}, we have
\be f_{D}(a,b) > 0 \quad . \label{fabm0} \ee
\no This subset of models are the ones we call TDiff henceforth. 
For arbitrary $a,b\in\mathbb{R}$, the model (\ref{L2}) may be identified with the linearized version of the following Lagrangian

\be \Lag = \sqrt{-g} \left\lbrack P(-g)R(g_{\mu\nu}) + Q(-g) g^{\mu\nu}\p_{\mu}g\p_{\nu}g \right\rbrack \quad . \label{lpq} \ee

\no where $P(-g)$ and  $Q(-g) $ are analytic functions at $g=-1$ such that $P(1)>0$, they are otherwise arbitrary. The linearization is obtained via $ g_{\mu\nu} = \eta_{\mu\nu} + h_{\mu\nu}/\sqrt{P(1)}$ with the identification $a=1+2\, P^{\prime}(1)/P(1)$, $b=1+4\, P^{\prime}(1)/P(1)+4\, Q(1)/P(1)$. Here we are mainly interested in massive theories obtained from \eqref{L2} via dimensional reduction so we are not going to explore \eqref{lpq} anymore, see e.g. \cite{bsz} for further developments and phenomenological applications when compared to general relativity.

It is known that TDiff is the minimal symmetry to have a Lorentz invariant S-matrix describing scattering of massless spin-2 particles \cite{van,herrero}. It is also the maximal subgroup of the diffeomorphisms.   We point out that   we can always treat TDiff as an one-parameter Lagrangian since, without loss of generality, we can use the $\lambda$-shift  and shift  $a\neq\frac{2}{D}\rightarrow a=1$ while  keeping $b$ the only free parameter. Similarly, we can fix $a=\frac{2}{D}$ and let $b$ free,  always bearing in mind that $f_{D}(a,b) > 0$ in both cases. The case $a=1$ may not be shifted into $a=2/D$ since this would require a non invertible $\lambda$-shift ($\lambda = -1/D$).

The Kaluza-Klein (KK) dimensional reduction of both Diff and  more recently WTDiff, see  \cite{bfh}, have been already discussed in the literature, their results will be compared to ours whenever possible as will be explained in the next section.

\section{Dimensional reduction}

\subsection{Particle content and unitary gauges}

In this subsection, we present the Kaluza-Klein (KK) dimensional reduction of the TDiff model and comment on different ways of gauging the Stückelberg fields away. First, we write the TDiff Lagrangian in $D+1$ dimensions, below $A,M,N=0,1,\cdots,D$,
\be
\Lag_{D+1}=-\frac{1}{4}\del_A H^{MN}\del^A H_{MN}+\frac{1}{2}\del_A H^{AM}\del_N H^{N}{}_{M}-\frac{a}{2}\del_M H^{MN}\del_N H+\frac{b}{4}\del_A H \del^A H , \label{LTD}
\ee
It is invariant under $\delta H_{MN}=\del_M \xi_N^T+\del_N \xi_M^T$ with 

\be \del^M \xi_M^T=0 \quad . \label{tcD+1} \ee

\no Notice that the requirement of physical particles in $D+1$ dimensions means\footnote{The  case $f_{D+1}=0$ contains the WTDiff and Diff theories in $D+1$.} 
\be f_{D+1}(a,b) = 1-2a-b(D-1)+a^2 D \ge 0 \label{D0}.\ee
Defining the cyclic coordinate $x_{D}=y$, we suitably decompose $H_{MN}(x,y)$:
\bea
&H_{\mu\nu}(x,y)= \sqrt{\frac{m}{\pi}} h_{\mu\nu}(x)\cos{my}, \label{huv} \\
&H_{y\mu}(x,y)=\sqrt{\frac{m}{\pi}}A_\mu(x)\sin{my}, \label{huy} \\
&H_{yy}(x,y)=\sqrt{\frac{m}{\pi}}\varphi(x)\cos{my}.
\label{hyy}
\eea
 The action in $D+1$ dimensions is
\be S = \int d^{D+1} x \; \Lag_{D+1} =  \int d^D x \; \int_0^{\frac{2\pi}{m}} dy \, \Lag_{(D+1)}.
\ee
Integrating over y, we obtain the massive theory in $D$ dimensions:
\bea
\Lag_D=&\Lag(a,b)-\frac{1}{4}F_{\mu\nu}^2 [A_\mu]-\frac{m^2}{4}(h_{\mu\nu}h^{\mu\nu}-b\;h^2)+\frac{a}{2}h_{\mu\nu}\del^\mu\del^\nu\varphi-m\;h_{\mu\nu}\del^\mu A^\nu+m\;a\;h\del_\mu A^\mu + \nn \\
&-m(a-1)A_\mu\del^\mu \varphi-\frac{1}{2}h[b\;\square+m^2(a-b)]\varphi-\frac{1}{4}\varphi [(b-1)\square+m^2(2a-b-1)]\varphi, \label{3.14}
\eea
where $\Lag(a,b)$ is the massless TDiff model in $D$ dimensions given in \eqref{L2}. Once again we split the discussion in three cases. 

\subsubsection{Massive Fierz-Pauli}

If $(a,b)=(1,1)$, the $D+1$ theory \eqref{LTD} is invariant under arbitrary diffeomorphisms without the restriction \eqref{tcD+1}. Consequently, \eqref{3.14} becomes invariant under full Diff and $U(1)$ transformations. We redefine the vector field in order to write down the gauge transformations in a diagonal form, i.e., 

\bea
\delta h_{\mu\nu} (x) &=& \del_\mu\psi_\nu + \del_\nu\psi_\mu,
\label{gaugehuv} \\
\delta a_\mu (x) &\equiv & \delta [A_{\mu} - \del_\mu \varphi/(2\,m)] = -m\;\psi_\mu ,
\label{gaugehuy} \\
\delta \varphi(x)&=& 2m\;\Lambda .
\label{gaugehyy}
\eea
where $\psi_{\mu}(x)$ and $\Lambda(x)$ are $D+1$ independent gauge parameters which stem from
\bea
&\xi_\mu(x,y)=\sqrt{\frac{m}{\pi}}\psi_\mu (x) \cos{my}, \label{psiu} \\
&\xi_y(x,y)=\sqrt{\frac{m}{\pi}}\Lambda (x) \sin{my}.
\label{psi}
\eea
The Lagrangian \eqref{3.14} becomes the usual massive spin-2 FP model
in $D$ dimensions written in terms of a gauge invariant field $H_{\mu\nu}^{Diff}$,

\bea \Lag_D(\HD ) &=& \Lag_{EH}(\HD) -\frac{m^2}4 \left\lbrack (\HD)^2-(H^{Diff})^2\right\rbrack =  \Lag_{FP}(\HD) \label{fp} \\
\HD &\equiv & h_{\mu\nu} + \frac{ \p_{\mu}a_{\nu}+ \p_{\nu}a_{\mu}}m = h_{\mu\nu} + \frac{ \p_{\mu}A_{\nu}+ \p_{\nu}A_{\mu}}m - \frac{\p_{\mu}\p_{\nu}\varphi}{m^2} \label{hd} \eea 
where $\Lag_{EH}$ is the linearized Einstein-Hilbert Lagrangian.

It is clear from \eqref{gaugehuv}-\eqref{gaugehyy} that $a_{\mu}$ and $\varphi$ are pure gauge. The massive spin-2 content of \eqref{3.14}, assured by the corresponding FP conditions $\p^{\mu}h_{\mu\nu}=0=h$ and the Klein-Gordon (KG) equation $(\Box - m^2)h_{\mu\nu}=0$, follows from the unitary gauge $a_{\mu}=0=\varphi$ which can be set at action level without affecting the particle content of the model. The gauge completely (uniquely) determine the gauge parameters $\psi_{\mu}$ and $\Lambda$, thus satisfying the ``completeness'' criterium of \cite{moto} for a good gauge to be fixed at action level. This is a key issue for the elimination of the Stückelberg fields as we will see in the next subsection.

\subsubsection{Massive WTDiff}

In the case $[a,b]=[2/(D+1),(D+3)/(D+1)^2]$, the model \eqref{LTD} becomes  the massless WTDiff model in $D+1$. Such dimensional reduction has been investigated in \cite{bfh}. Our Lagrangian \eqref{3.14} coincides with 
the corresponding one of \cite{bfh} in the above case. Although it stems from a WTDiff model in $D+1$, \eqref{3.14} is invariant under full Diff and Weyl transformations. In order to single out the pure gauge degrees of freedom we find convenient to redefine the vector and scalar fields 

\bea a_{\mu}^W &=& A_{\mu} + \frac{\p_{\mu}(h-\, D\, \varphi )}{2(D+1)\, m}, \label{amw} \\
\Phi &=& \varphi + h.
\eea
Accordingly, we have the following  Diff  and Weyl  transformations:

\bea
\delta h_{\mu\nu}  &=& \del_\mu\psi_\nu + \del_\nu\psi_\mu + \eta_{\mu\nu} \, \chi
\label{dhmn} \\
\delta a_\mu^W  &=&  -m\;\psi_\mu ,
\label{damw} \\
\delta \Phi &=& (D+1)\chi ,
\label{dbigphiw}
\eea
Notice that the $U(1)$ parameter $\Lambda$ has been eliminated via the constraint 
 
\be 
\del^\mu\psi_\mu+m\,\Lambda=0.
\label{vinculo}
\ee
which follows from the higher dimensional transverse condition \eqref{tcD+1}.  The previous gauge transformations suggest the definition of a gauge invariant tensor as in the Diff case, namely, 

\be \HWT = h_{\mu\nu} + \frac{\p_{\mu}a_{\nu}^W + \p_{\nu}a_{\mu}^W}{m} - \frac{\eta_{\mu\nu}\Phi}{D+1}  \label{hwt} \ee

\no It turns out that the whole Lagrangian \eqref{3.14}  reduces to the FP model:

\be \Lag_D = \Lag_{FP}(\HWT)= \Lag_{FP}\left(h_{\mu\nu} + \frac{\p_{\mu}a_{\nu}^W + \p_{\nu}a_{\mu}^W}{m} - \frac{\eta_{\mu\nu}\Phi}{D+1} \right) \label{ldwt} \ee

\no Notice that the replacement $h_{\mu\nu} \to \HWT $ is non trivial since it envolves double derivaties of the tensor field itself. However, although the $U(1)$ symmetry disappears, we can use  the Weyl symmetry altogether with the Diff symmetry in order to eliminate pure gauge degrees of freedom and fix the unitary gauge $a_{\mu}^W=0=\Phi$. Therefore, we recover spin-2 massive particles in $D$ dimensions which is the expected result for a dimensional reduction of spin-2 massless particles in $D+1$. The authors of \cite{bfh} have chosen the partial gauge fixing $a_{\mu}^W=0$ which leaves us with a massive model with Weyl symmetry. It amounts to the FP model with the replacement $h_{\mu\nu} \to h_{\mu\nu} -\eta_{\mu\nu}h/D + \eta_{\mu\nu}\phi $ which comes from \eqref{ldwt} after an invertible redefinition of the scalar field $\Phi= (D+1)(h/D-\phi)$ where $\phi$ is now Weyl invariant. We stress that such partial gauge does satisfy the completeness criterion of \cite{moto}, since $a_{\mu}^W=0$ completely (uniquely) fix the $D$ parameters $\psi_{\mu}$, consequently it can be fixed at action level without problems even though the Weyl symmetry is left unbroken.

We have found interesting from the point of view of the completeness criterium of \cite{moto} to discuss a third gauge condition:

\be f_{\mu}\equiv A_{\mu}- \frac{c}{m} \p_{\mu}h =0 \quad ; \quad \varphi =0 \quad . \label{gc3} \ee
where $c$ is so far an arbitrary real constant. Are we allowed to fix such gauge at action level? After this gauge is fixed, the action \eqref{ldwt} only depends upon the tensor field. We know that the only viable choice for massive spin-2 particles is the FP model or at most a $\lambda$-shifted ($h_{\mu\nu}\to h_{\mu\nu} + \lambda \eta_{\mu\nu}\, h$) version thereof. Indeed, under gauge transformations the gauge conditions change as

\bea \delta\varphi &=& \chi -2\,\p_{\mu}\psi^{\mu} \label{gt1} \\
\delta f_{\mu} &=& -m\, \psi_{\mu}-\frac{(1+2\, c)}m \p_{\mu}(\p \cdot \psi)- \frac{c\, D}m\p_{\mu}\chi \label{gt2} \eea
Under the residual transformations

\be \delta_{\gamma} \psi_{\mu} = \p_{\mu}\gamma \quad ; \quad \delta_{\gamma}\chi=2\,\Box \gamma \label{gamma} \ee
we have 

\be \delta_{\gamma}\delta\varphi = 0 \quad ; \quad
\delta_{\gamma}\delta f_{\mu} = - \p_{\mu}\left\lbrack m\gamma + [1+ 2\, c(1+D)]\frac{\Box \gamma}m\right\rbrack \quad  \label{dd} \quad . \ee

\no Therefore, if $c\ne -1/(2(1+D))$ the above transformations tell us that 
the gauge conditions \eqref{gc3} do not completely (uniquely) fix the gauge parameters $\psi_{\mu}$ and $\chi$, since we can always add the $\gamma$-transformations \eqref{gamma} such that $\gamma$ satisfies the Klein-Gordon equation 
$\Box \gamma = - m^2\,\gamma/\left(1+ 2\, c(1+D)\right)$. On the other hand, if 
$c = -\frac{1}{2(1+D)}$ the gauge conditions will completely fix the gauge parameters and  consequently we are allowed, according to \cite{moto}, to fix the third gauge at action level. In fact, it is only at that particular point that the gauge fixed version of \eqref{ldwt} becomes a $\lambda$-shifted version of the FP massive theory, namely, $\Lag_{FP}[h_{\mu\nu}-\eta_{\mu\nu}h/(D+1)]$. For any $c\ne -\frac{1}{2(1+D)}$ the action \eqref{ldwt} at the gauge \eqref{gc3} is  not unitary \cite{fp,pvn73}. 

\subsubsection{Massive TDiff model}

This is the most involved case, the constants $(a,b)$ are such that $f_{D+1}(a,b)>0$. Differently from the previous two cases, the particle content of \eqref{LTD} consists of a massless spin-2 and a massless spin-0 particle. Thus, we expect massive spin-2 and massive spin-0 fields in $D$ dimensions as will turn out to be the case. Once again the $U(1)$ symmetry is eliminated via 
\eqref{vinculo}. So we have one less symmetry than in the previous two cases. We are left only with full Diff. Redefining the vector field as in the Diff case and the scalar field as in the WTDiff case we have

\bea
\delta h_{\mu\nu} (x) &=& \del_\mu\psi_\nu + \del_\nu\psi_\mu,
\label{deltah} \\
\delta a_\mu (x) &\equiv & \delta [A_{\mu} - \p_{\mu} \varphi/(2\,m)] = -m\;\psi_\mu ,
\label{deltaa} \\
\delta \Phi &=& \delta (\varphi+h)= 0.
\label{deltaphi}
\eea
After trying an Ansatz for $\HTD$ and for a massive scalar-tensor theory, we have been able to show that \eqref{3.14} can be written as

\be \Lag_D = \Lag_{FP}(\HTD) + \frac{f_{D+1}(a,b)}{4(D-1)} \,\Phi\left(\Box - m^2 \right)\Phi \label{ldtd} \ee
where we have defined the gauge invariant tensor

\be \HTD = h_{\mu\nu} + \frac{\p_{\mu}a_{\nu} + \p_{\nu}a_{\mu}}{m} + \frac{a-1}{D-1}\left(\eta_{\mu\nu} - \frac{\p_{\mu}\p_{\nu}}{m^2}\right)\Phi  \label{htd} \ee

Notice that \eqref{ldtd} and \eqref{htd} include the two previous cases which satisfy $f_{D+1}(a,b)=0$. The unitary gauge now corresponds to
$ a_{\mu} + (1-a)\p_{\mu}\Phi/[ 2\, m\, (D-1)]=0 $, after which we redefine $h_{\mu\nu} \to h_{\mu\nu} -\frac{a-1}{D-1}\eta_{\mu\nu}\Phi$ and decouple the massive spin-2 field from the scalar one. Remarkably, unitarity, $f_{D+1}(a,b)>0$,  of the $D+1$ model \eqref{LTD} is strictly preserved by the dimensional reduction. In section 4 we find interesting to choose the gauge $a_{\mu}=0$. This and the unitary gauge can be fixed at action level without changing the content of the theory. Before we leave this section, we look at the massless limit of \eqref{3.14}.

\subsection{Smooth massless limit}

The reduced theory \eqref{3.14} at $m=0$ becomes

\be
\Lag_D^{m=0}= \Lag(a,b)+\frac{a}{2}h_{\mu\nu}\del^\mu\del^\nu\varphi-\frac{b}{2}h\, \square \varphi+\frac{1-b}{4}\varphi\square\,\varphi -\frac{1}{4}F_{\mu\nu}^2 [A_\mu] \quad .  \label{3.15}
\ee
The Lagrangian \eqref{3.15} is invariant under TDiff $\delta h_{\mu\nu} = \p_{\mu}\psi_{\nu}^T + \p_{\nu}\psi_{\mu}^T$ and $U(1)$, $\delta A_{\mu} = \p_{\mu}\Lambda$. If $a=1=b$, TDiff is enlarged to full Diff while at $[a,b]=[2/(D+1),(D+3)/(D+1)^2]$ is enlarged to WTDiff where the scalar field must contribute to the Weyl symmetry: $(\delta h_{\mu\nu},\delta\varphi) = (\eta_{\mu\nu}\chi , \chi)$. In this sense the model \eqref{3.14} is a massive deformation of a Diff, WTDiff and TDiff theory in the corresponding three cases respectively. Regarding the particle content of \eqref{3.15}, we have to split it in two cases. First, if $a\ne 2/D$,  after a field redefinition $h_{\mu\nu} \to  h_{\mu\nu} + \frac{a}{2-a\, D}\eta_{\mu\nu}\varphi$, we have:

\be
\Lag_D^{m=0}= \Lag (a,b) + Y \, h\, \Box \, \varphi + \frac{Z}2\varphi \, \Box \, \varphi -\frac{1}{4}F_{\mu\nu}^2  \quad . \label{lyz} \ee

\be Y = \frac{a+a^2-b}{2(2-a\, D)} \quad ; \quad Z = \frac{(D-2)\left\lbrack a(D+1)-2\right\rbrack^2+4\, f_{D+1}(a,b)}{2(D-1)(2-a\, D)^2}
\quad . \label{yz} \ee

In the case of the reduction of the   $D+1$ WTDiff model  $[a,b]=[2/(D+1),(D+3)/(D+1)^2]$ we have $Y=0=Z$. So we end up with a continuous massless limit with (in $D=4$) $3+2= 2s+1$ physical degrees of freedom, since the reduced Lagrangian and the unitarity condition for the TDiff model in $D$ dimensions become respectively

\be
\Lag_D^{m=0}= \Lag\left(\frac{2}{D+1},\frac{D+3}{(D+1)^2}\right) -\frac{1}{4}F_{\mu\nu}^2  \label{lwtm0} \ee

\be f_{D}\left(\frac{2}{D+1},\frac{D+3}{(D+1)^2}\right)=\frac{D-1}{(D+1)^2} > 0 \quad . \label{fdwt} \ee

 In the Diff case $a=1=b$, since  $Y=0$, $Z>0$ and  $\Lag(1,1)$ is the usual EH Lagrangian, we have once again a smooth massless limit with $2+2+1= 2s+1$ physical degrees of freedom in $D=4$.

In the pure TDiff case, still assuming $a\ne 2/D$, after the redefinition $\varphi \to \varphi - Y\, h/Z$ in \eqref{lyz} we have a TDiff model plus Maxwell and a decoupled scalar field:

\be
\Lag_D^{m=0}= \Lag(a,\tilde{b})  -\frac{1}{4}F_{\mu\nu}^2  + \frac{Z}2 \varphi \Box \,\varphi \label{ltd} \ee

\no where $Z$ is given in \eqref{yz} and 

\bea \tilde{b} &=& b + \frac{(D-1)(a^2+a-2b)^2}{(D-2)[a(D+1)-2]^2 + 4 f_{D+1}(a,b)} \quad , \label{btil} \\
f_D(a,\tilde{b}) &=& \frac{(D-1)(2-a\, D)^2f_{D+1}(a,b)}{(D-2)[a(D+1)-2]^2 + 4 f_{D+1}(a,b)} \quad . \label{fdabtil} \eea

\no 
Due to $f_{D+1}(a,b)>0$ it follows that $Z$ and $f_D(a,\tilde{b})$ are both positive and we have again a smooth massless limit with 
$3+2+1=2s+1 +1$  physical degrees of freedom in $D=4$. Finally, if $a=2/D$, after $\varphi \to \varphi - h$ in \eqref{3.15} followed by 
$h_{\mu\nu}\to h_{\mu\nu} +\eta_{\mu\nu}\varphi/D$ and another change
$\varphi \to \varphi -h/(2\, D\, \tilde{Z})$ we obtain,

\be
\Lag_D^{m=0}= \Lag(0,\tilde{B})  -\frac{1}{4}F_{\mu\nu}^2  + \frac{\tilde{Z}}2 \varphi \Box \,\varphi \label{ltd2d} \ee
where

\be \tilde{B} = \frac{D+1-b\, D^2}{2\, D^2\, \tilde{Z}} \quad ; \quad \tilde{Z} = \frac{D-2+D^2f_{D+1}(2/D,b)}{2\, D^2(D-1)} \quad . \label{btilz} \ee 

\no The unitarity condition of the $D+1$ TDiff model 
 $f_{D+1}(2/D,b) >0$ assures $\tilde{Z}>0$ and unitarity of the TDiff model, since  

\be f_{D}(0,\tilde{B}) = \frac{D^2(D-1)f_{D+1}(2/D,b)}{D-2+D^2 f_{D+1}(2/D,b) } > 0 \quad . \label{fdbtil}\ee

\no Thus, we have a smooth massless limit in all cases.

\section{Massive reduced model at $A_{\mu} = \p_{\mu} \varphi/(2\,m)$}

\subsection{General remarks}

 The model \eqref{ldtd} at the gauge $ a_\mu = A_{\mu} - \frac{\p_{\mu} \varphi}{2\,m}=0$  becomes the massive scalar tensor theory:

\bea \Lagstm &=& \Lag (a,b)-\frac{m^2}{4}(h_{\mu\nu}h^{\mu\nu}-b\;h^2)+\frac{(a-1)}{2}h_{\mu\nu}\del^\mu\del^\nu\varphi+\frac{(a-b)}{2}h(\square-m^2)\varphi \nn\\ &+&\frac{(2a-b-1)}{4}\varphi(\square-m^2)\varphi \quad . \label{ltdm} 
\eea

\no The first remark we make is that in both cases of Diff
 and TDiff, see \eqref{gaugehuy} and \eqref{deltaa},  we have $\delta a_{\mu} = -m\,\psi_{\mu}$. Thus, the gauge $a_{\mu}=0$ completely determines the parameters $\psi_{\mu}$. However, from \eqref{amw}-\eqref{dbigphiw} we see that in the WTDiff case $\delta a_{\mu} = -m\, \psi_{\mu}- \p_{\mu}\chi/(2m)$ which has the residual symmetry $(\delta\psi_{\mu},\delta\chi)= (- \p_{\mu}\gamma,2\, m^2 \gamma)$. Therefore,  the gauge $a_{\mu}=0$ at action level is not allowed at the WTDiff point $(a,b)=\left(\frac{2}{D+1},\frac{D+3}{(D+1)^2}\right)$. So henceforth our results do not apply in this special case and may not be compared with \cite{bfh} any more.

Our second remark concerns the massless limit of \eqref{ltdm}, i.e.,
\be 
\Lagsto=\Lag (a,b)+\frac{a-1}{2}\varphi\del^\mu\del^\nu h_{\mu\nu}+\frac{a-b}{2}\varphi\square h+\frac{2a-b-1}{4}\varphi\square\varphi,
\label{lm0}
\ee
\no Although a scalar-tensor Lagrangian of the type $\Lag (a,b)+c_1 \varphi \del_\mu\del_\nu h^{\mu\nu} + c_2 \varphi \square h + c_3 \varphi \square \varphi $ is in general only invariant under TDiff, the coefficients in \eqref{lm0} are such that we have full diffeomorphism invariance:

\be 
\delta h_{\mu\nu}=\del_\mu\psi_\nu+\del_\nu\psi_\mu,\qquad \delta\varphi=-2\nabla\cdot\psi,
\label{simetriam0}
\ee
Thus, \eqref{ltdm} is a massive deformation of Diff instead of TDiff! The same situation ocurred in \cite{bfh},  where the dimensional reduction of a WTDiff model has given rise  to a Weyl Stückelberg version $(h_{\mu\nu}\rightarrow h_{\mu\nu}-\eta_{\mu\nu}h/D + \varphi \eta_{\mu\nu})$ of the usual massive Fierz-Pauli (FP) model whose mass terms break Diff instead of TDiff. It seems that massive spin-2 particles require the breakdown of full diffeomorphisms in order to produce the correct number of constraints. In \cite{blas} it has been shown that there is no Lorentz covariant mass term that could be added to the TDiff pure tensor model \eqref{L2} that might generate a stable theory of massive spin-2 particles. It seems that the addition of a scalar field does not help either. In session 5 we show that the Diff symmetry is required in the flat limit in order to have a vector constraint in a massive scalar tensor theory of second order in derivatives. The reader may find our conclusion doubtful from the point of view of the massless limit of \eqref{vinculo}, but notice that our gauge, and the vector gauge used in \cite{bfh},  is singular at $m\to 0$.

 Notwithstanding, it is not fully inappropriate to call \eqref{ltdm} a massive TDiff model since the particle content of \eqref{lm0} is exactly the same of a TDiff model in $D$ dimensions, see \eqref{decoup}, namely a physical massless spin-2 particle plus a massless spin-0 particle which is unitary whenever $f_D(a,b)>0$. If we compare \eqref{lm0} to \eqref{L2} we have one more field but one more symmetry, longitudinal diffeomorphisms (LDIFF): $\delta h_{\mu\nu} = \p_{\mu}\p_{\nu}\lambda $. However, the equivalence between \eqref{lm0} and \eqref{L2} is not complete. It depends  on the boundary conditions. In fact, \eqref{lm0} stands to \eqref{L2} as LEH stands to the WTDiff model. Namely,
 if we first derive the equations of motion $E_{\mu\nu}=\delta S^{ST}_0/\delta h^{\mu\nu} = 0$ and $\delta S^{ST}_0/\delta \varphi = 0$ and then fix the gauge condition $\varphi =0$ we have $R(a,b) \equiv  (a-1)\p^{\mu}\p^{\nu}h_{\mu\nu} +(a-b) \Box \, h = 0$. On the other hand, if we first fix  $\varphi =0$ at action level, we lose its equation of motion and we can only derive from $\p^{\mu}E_{\mu\nu}=0$ that $R(a,b)=c$ where $c$ is a real constant not necessarily zero. We have lost information since the gauge $\varphi =0$ does not completely determine the gauge parameters $\psi_{\mu}$ due to the residual symmetry 
 $\delta \psi_{\mu} = \p_{\mu} \gamma $ with $\Box \gamma =0$.
 This is similar to the fact that $h=0$ at action level in the LEH action make us lose the trace of the linearized Einstein-Hilbert equation which becomes an integration constant related to the cosmological constant in the WTDiff model. So TDiff and WTDiff are obtained respectively from $\Lag_0^{ST}$  and $\Lag_{EH}$ via an ``illegal'' gauge condition  at action level. A stronger claim at nonlinear level, see \cite{villarejo},  is that transverseDiff gravity \eqref{lpq} is to scalar tensor as unimodular gravity is to general relativity.

\subsection{Helicity variables and equations of motion}

In this subsection we use helicity variables, see e.g. \cite{deserprl}, in order to identify the different helicity modes of the massive spin-2 particle and split them from the scalar field in \eqref{ltdm} without introducing field redefinitions involving time derivatives which might spoil the canonical structure of the theory. The helicity decomposition is also called sometimes ``the cosmological decomposition'' \cite{blas,mukhanov}. The symmetric tensor $h_{\mu\nu}$ is decomposed in its scalar, vector and purely tensor modes, similarly to \cite{blas},  
\bea
&h_{00}=A, \qquad h_{0i}=\del_i B+V^T_i, \label{h00} \\
&h_{ij}=\psi \delta_{ij}+\omega_{ij}E+2\del_{(i}F^T_{j)}+h^{TT}_{ij}, \label{hij}
\eea
where $\omega_{ij}=\p_i\p_j/\nabla^2 $, while $V^T_i$ and $F^T_i$ are transverse vectors, e.g. $\p_jV^T_j=0$. The tensor $h^{TT}_{ij}$ is  traceless and transverse. Applying to the Lagrangian \eqref{ltdm}, and eliminating the fields $A,B,E$ and $V^T_i$ via their equations of motion, saving many details specially in the messy scalar sector, we end up with
%\footnote{See \cite{masterrr} for a positivity proof of the reduced Hamiltonian and a complete constraints analysis based on the Dirac algorithm.} 
\be 
\Lag_m= \Lag_m^t+\Lag_m^v+\Lag_m^s,
\ee  
where
\be 
\Lag_m^t=\frac{1}{4}h^{TT}_{ij}(\square-m^2)h^{TT}_{ij} \quad ; \quad \Lag_m^v=\frac{1}{2}\tilde{F}^T_i(\square-m^2)\tilde{F}^T_i \quad ,
\ee
\be 
\Lag_m^s=\frac{(D-1)(D-2)}{4}\theta(\square-m^2)\theta+\frac{f_{D+1}(a,b)}{4(D-1)}\Phi(\square-m^2)\Phi, \label{scalardec}
\ee
Therefore, $h^{TT}_{ij}$, $\tilde{F}^T_i\equiv \sqrt{\frac{-\nabla^2}{m^2-\nabla^2}}m F_i^T$ and  $\theta\equiv \psi+\frac{a-1}{(D-1)}\Phi$, represent the $\pm 2,\pm1$ and $0$ helicity modes of the spin-2 sector respectively, while $\Phi = \varphi + h $ stands for the spin-0 scalar particle whose propagation is physical if $f_{D+1}(a,b)\ge 0 $. This is exactly the unitarity condition for the massless TDiff theory in $D+1$ dimensions which is the origin of \eqref{ltdm}.
This completes the analysis of the particle content of \eqref{ltdm} 
which coincides with the content of \eqref{ldtd} at the unitary gauge mentioned in subsection 2.1. It is thus established  that the gauge $A_{\mu}=\p_{\mu}\varphi/(2m)$ can be fixed at action level without spoiling  the physical content of the reduced model.

For future generalizations to curved backgrounds we find instructive to look at the equations of motion of \eqref{ltdm}. Before we do it, however, let us make the redefinition inspired by \eqref{ldtd}, $\varphi= \Phi - h $. The tensor and scalar equations of motion become respectively,
\bea
E_{\mu\nu}\equiv (\square-m^2) h_{\mu\nu}-\del^\alpha(\del_\mu h_{\nu\alpha}+\del_\nu h_{\mu\alpha}) + \del_\mu \del_\nu h + \eta_{\mu\nu} \del^{\alpha}\del^\beta h_{\alpha\beta}-\eta_{\mu\nu}(\square-m^2)h\nn \\ +(a-1)\del_\mu\del_\nu\Phi+(1-a)(\square-m^2)\eta_{\mu\nu}\Phi=0, \label{tensoeq}
\eea
\be 
\Psi \equiv (a-1)\del^\mu\del^\nu h_{\mu\nu}+(1-a)(\square-m^2)h+(2a-b-1)(\square-m^2)\Phi=0. \label{Phieq}
\ee
From \eqref{tensoeq} we have the vector constraint :
\be 
\del^\mu E_{\mu\nu}=-m^2\del_\mu h^{\mu\nu}+m^2\del_\nu h+(a-1)m^2\del_\nu\Phi=0. \label{vecto}
\ee 
In the Diff case $a=1=b$ we get rid of the scalar field and
the following combination provides a scalar constraint $m^2\eta^{\mu\nu}E_{\mu\nu}+(D-2)\del^\mu\del^\nu E_{\mu\nu}=m^2(D-1)h=0$, back in \eqref{tensoeq}  and \eqref{vecto}  we obtain the
usual FP conditions $h=0=\p^{\mu}h_{\mu\nu}$ and $(\Box -m^2)h_{\mu\nu}=0$. Henceforth we assume the pure TDiff case $f_{D+1}(a,b)>0$.
The following combination supplies a scalar constraint
\bea 
\Omega&=&(a^2-b)m^2\eta^{\mu\nu}E_{\mu\nu}+f_{D}(a,b)\del^\mu\del^\nu E_{\mu\nu}+(a-1)m^2\Psi =0 \nn \\
&=& m^2f_{D}(a,b)\left[h+(a-1)\Phi\right] = 0 \Rightarrow \boxed{h+(a-1)\Phi = 0} \quad, \label{hphi}
\eea
provided that $f_{D}(a,b)>0$ which follows from $f_{D+1}(a,b)>0$ via the identity $(D-1)f_{D}(a,b)=(a-1)^2 +(D-2)f_{D+1}(a,b)$. From \eqref{hphi} in \eqref{vecto} we have $\del^\mu h_{\mu\nu}=0$. Additionally, \eqref{tensoeq} reads simply $(\square-m^2)h_{\mu\nu}=0$, back in \eqref{Phieq} we have the spin-0 Klein-Gordon equation

\be 
(\square-m^2)\Phi=0.
\ee
A traceless and transverse tensor can be easily built
\be 
H_{\mu\nu}=h_{\mu\nu}-\frac{1}{D-1}\left(\eta_{\mu\nu}-\frac{\del_\mu\del_\nu}{m^2}\right)h \quad , \label{Hmn}
\ee
and shown to satisfy the conditions for massive spin-2 particles,
\be 
\begin{cases}
(\square-m^2)H_{\mu\nu}=0,\\
\del^\mu H_{\mu\nu}=0,\\
\eta^{\mu\nu}H_{\mu\nu}=0,
\end{cases}
\ee
Notice that, due to \eqref{hphi}, \eqref{Hmn} is nothing but
$\HTD$, see \eqref{htd}, at the gauge $a_{\mu}=0$.

\subsection{Massive scalar-tensor coupled to sources}

Now we couple \eqref{ltdm} to arbitary tensor and scalar sources
and investigate the influence of the scalar field
in the vDVZ mass discontinuity \cite{vdvz}. The arbitrariness of the sources allows us to carry out an off-shell Lorentz covariant proof of unitarity.  Motivated by simplicity and the previous discussions about the separation of the spin-2 and spin-0 degrees of freedom let us redefine $\varphi \to \Phi - h$ and  $h_{\mu\nu}\rightarrow h_{\mu\nu}+\frac{1-a}{D-1}\eta_{\mu\nu}\Phi$ before adding sources. This cancels out the scalar tensor coupling in the mass term  ($m^2 h\Phi$) and leads to the FP Lagrangian in the tensor sector. Adding sources we have
\be
\Lag_m (T,J)=\Lag_{FP}+\frac{a-1}{2(D-1)}\Phi(\del^\mu\del^\nu h_{\mu\nu}-\square h)+\frac{f_{D+1}}{4(D-1)}\Phi (\square-m^2)\Phi + h_{\mu\nu}T^{\mu\nu}+\Phi J. \label{lmtj}
\ee
Since there is no local symmetry any more, the sources $T_{\mu\nu}$ and $J$ are totally arbitrary.  Integrating over $h_{\mu\nu}$ and $\Phi$ in the path integral we derive

\be {\cal Z}[T,J] = \exp \left\lbrace-\imath \int\, d^Dx\, d^D y \left\lbrack T_{\mu\nu}(x)\left(G_{FP}^{-1}(x,y)\right)^{\mu\nu\alpha\beta}T_{\alpha\beta}(y) + {\cal J}(x)G_S^{-1}(x,y){\cal J}(y)\right\rbrack\right\rbrace, \label{ztj}\ee
where, suppressing the indices, 

\bea G^{-1}_{FP} &=& \frac{P_{ss}^{(2)}}{(\square-m^2)}-\frac{P_{ss}^{(1)}}{m^2}+\frac{(2-D)(\square-m^2)P_{ww}^{(0)}}{m^4(1-D)}+\frac{P_{sw}^{(0)}+P_{ws}^{(0)}}{m^2\sqrt{D-1}}, \label{gfp} \\
 G^{-1}_{S} &=&   \frac{D-1}{f_{D+1}}\frac{1}{(\square-m^2)}, \label{gs} \\ \mathcal{J}&=& J+\frac{(a-1)}{(D-1)m^2}\p^{\mu}\p^{\nu}T_{\mu\nu} \label{j} \eea
The projection and transition operators $P_{IJ}^{(s)}$, where $s$ stands for the spin of the projected subspace, are given in the appendix. The two-point amplitude in momentum space is given by
\be 
{A}_{ST}(p) =  -\imath\, \left\lbrack T^*_{\mu\nu}(p)\left( G^{-1}_{FP}(p)\right)^{\mu\nu\alpha\beta}T_{\alpha\beta}(p) + {\cal J}^*(p)\left( G^{-1}_{S}(p)\right){\cal J}(p) \right\rbrack .
\ee
\no The imaginary part of the residue of ${A}_{ST}(p)$ at the massive pole is given by 
\be 
Im Res(\mathcal{A}_{ST})_{|p^2\rightarrow-m^2}= T^{*} P_{ss}^{(2)} T (p) + \frac{D-1}{f_{D+1}(a,b)} |\mathcal{J}|^2,
\ee
which is definitely positive, if $f_{D+1}>0$, since the first term is nothing but the usual FP result known to be positive.  Consequently we have an off-shell Lorentz covariant proof of unitary
for the scalar tensor model  \eqref{ltdm}. Notice also that after replacing $J$ in (\ref{lmtj}) following (\ref{j}) and
taking into account the field redefiniton before (\ref{lmtj}), we see that the tensor field effectively coupled to $T^{\mu\nu}$ is $H_{\mu\nu}^{TD}$ given in \eqref{htd} at the gauge $a_{\mu}=0$.

Regarding the vDVZ mass discontinuity, if we assume that the scalar source is proportional to the trace of spin-2 source, ${\cal J}=c\, T$, with some real constant $c$, from \eqref{gfp} and \eqref{gs} we have the following tensor structure at $m\to 0$ in $D=4$:

\be G^{-1}_{\mu\nu\alpha\beta} = \frac{1}{\Box}\left\lbrace \frac{\eta_{\mu\alpha}\eta_{\nu\beta}+\eta_{\mu\beta}\eta_{\nu\alpha}}2 - \frac{\eta_{\mu\nu}\eta_{\alpha\beta}}2 + \eta_{\mu\nu}\eta_{\alpha\beta}\left\lbrack  \frac 16 + \frac{3\, c^2}{f_5(a,b)}\right\rbrack\right\rbrace + \cdots \label{gmenos1} \ee 
where dots stand for analytic contributions which only lead to contact terms in the two-point amplitude. If we neglect the two terms inside the brackets we have the Einstein-Hilbert result. The $1/6$ factor is the well known vDVZ mass discontinuity. Since $f_5(a,b)>0$, we see that the massive scalar field contribution is no cure for the mass discontinuity which is pushed even farther away.

%Finally, by using the definition of $\mathcal{J}$ and integrating by parts, we can rewrite the coupling to sources as
%\be
%h_{\mu\nu}T^{\mu\nu}+\phi J= \left(h_{\mu\nu}+\frac{1-a}{D-1}\frac{\del_\mu\del_\nu}{m^2}\phi\right)T^{\mu\nu}+\phi \mathcal{J} 
%\ee
%So, if we recall the tensor field redefinition made before \eqref{lmtj} we finally identify the field inside parenthesis with $\HTD$, defined in \eqref{htd}, as the pure spin-2 field.
%

\subsection{TDiff coupled to sources}

Now we return to the massless case and study the spin-0 and spin-2 contributions to scattering amplitudes in the pure tensor  TDiff model \eqref{L2}. We couple the massless TDiff model $\Lag(a,b)$ to sources, but from a slightly different perspective from \cite{blas}. First we remark that invariance of the source term $\int d^D x T_{\mu\nu}h^{\mu\nu}$ under TDiff does not require conserved sources, it only demands\footnote{See \cite{jps,astorga,ccg} for recent works with non conserved energy momentum tensors.} that 
\be 
\del_\mu T^{\mu\nu}=\del^\nu J,
\label{tdiffsource}
\ee 
where $J(x)$ is some local scalar quantity, which simply measures the non-conservation of $T_{\mu\nu}$. 

First of all, we have found interesting to rewrite the source term following a curious non local correspondence that we have found between $\Lag(a,b)$ and the massless theory $\Lagsto$ given in \eqref{lm0}.
Notice that after the invertible redefinition:
\be
h_{\mu\nu}=H_{\mu\nu}-\frac{(a-1)}{D-2}\eta_{\mu\nu}\Phi \quad ; \quad 
\varphi=-H+\frac{(a D-2)}{D-2}\Phi, \label{fr2}
\ee
the Lagrangian \eqref{lm0} decouples
\be 
\Lagsto=\Lag_{EH}(H_{\mu\nu})+\frac{f_{D}}{4(D-2)}\Phi\square\Phi,\label{decoup}
\ee
The fact that \eqref{decoup} has exactly the same particle content of $\Lag (a,b)$ in $D$ dimensions has inspired us to find a closer connection between those models. If  we substitute $h_{\mu\nu}\rightarrow h_{\mu\nu}+\frac{\del_\mu\del_\nu}{\square}\varphi$ in \eqref{L2}, we exactly reproduce \eqref{lm0}. The field $\varphi$ works like a Stückelberg field for longitudinal diffeomorphisms (LDiff). It leads to the symmetry $\delta h_{\mu\nu} = \del_\mu\del_\nu \Lambda$, $\delta \varphi = - \square \Lambda$, thus enlarging TDiff to Diff. By further taking into account \eqref{fr2}, we can go from \eqref{L2} directly to the decoupled scalar-tensor model \eqref{decoup} through the singular non invertible redefinition\footnote{There is also a non local field redefinition decoupling spin-2 and spin-0 fields in the appendix C of \cite{cf}, we thank A. Campoleoni and D. Francia for bringing  \cite{cf} to our knowledge.}
\be 
h_{\mu\nu}=H_{\mu\nu}-\frac{\del_\mu\del_\nu}{\square}H+\frac{(a D-2)}{D-2}\frac{\del_\mu\del_\nu}{\square}\Phi+\frac{(1-a)}{D-2}\eta_{\mu\nu}\Phi. \label{redeftdeh}
\ee
In particular, at the $D$ dimensional WTDiff point $(a,b)=\left(\frac{2}{D},\frac{D+2}{D^2}\right)$, where $f_{D}=0$, the non invertible redefinition\footnote{The last term $\eta_{\mu\nu}\Phi$ in \eqref{redeftdeh} drops out due to Weyl invariance.} $h_{\mu\nu}=H_{\mu\nu}-\frac{\del_\mu\del_\nu}{\square}H$ takes us from the WTDiff model to the linearized EH theory. The Weyl symmetry is broken and replaced by LDiff, such that WTDiff$\rightarrow$Diff.

Now we are ready to go back to $\Lag (a,b)$ with sources. If we substitute \eqref{redeftdeh} in the source term and use \eqref{tdiffsource}, we have (up to total derivatives)
\be 
h_{\mu\nu}T^{\mu\nu}=H_{\mu\nu}\tilde{T}^{\mu\nu}+\Phi \tilde{J}, \label{ht}
\ee
where
\be
\tilde{T}_{\mu\nu}=T_{\mu\nu}-\eta_{\mu\nu}J \quad ; \quad 
\tilde{J}=\frac{1-a}{D-2}T+\frac{aD-2}{D-2}J \label{tildetj}
\ee
The conservation of $\tilde{T}_{\mu\nu}$ follows from \eqref{tdiffsource}: 
\be 
\del^\mu\tilde{T}_{\mu\nu}=0.
\label{tdiffc}
\ee
Expressions \eqref{decoup} and \eqref{ht} suggest that $\tilde{T}_{\mu\nu}$ and $\tilde{J}$ are the right source combinations which couple to the spin-2 and spin-0 modes of the TDiff theory without mixing. Indeed, we can invert \eqref{tildetj} and obtain 
\be 
T_{\mu\nu}=\tilde{T}_{\mu\nu}+\eta_{\mu\nu}\left[\tilde{J}+\frac{(a-1)}{D-2}\tilde{T}\right]. \label{source}
\ee
Since \eqref{tildetj} is a fully invertible source redefinition, we interpret  \eqref{source} as a convenient way, without loss of generality, of rewriting  the original non conserved source in terms of a conserved one. Adding the source term and a gauge fixing one to $\Lag (a,b)$ , we have 
\be 
\Lag(a,b,T)=\frac{1}{2}h_{\mu\nu}\mathcal{O}^{\mu\nu,\alpha\beta}h_{\alpha\beta}+ h_{\mu\nu}T^{\mu\nu}. \label{lagt}
\ee
where, suppressing indices,
\bea
\mathcal{O}= \frac{\square}{2}P_{ss}^{(2)}+\frac{(1-b(D-1))\square}{2}P_{ss}^{(0)}+\frac{(2a-b-1)\square}{2}P_{ww}^{(0)}&+&\frac{\sqrt{D-1}(a-b)\square}{2}(P_{sw}^{(0)}+P_{ws}^{(0)})\nn\\ &-& \lambda \frac{\square^3}{2}P_{ss}^{(1)}. \label{ot}
\eea
The last term stands for the transverse gauge $\lambda(\del_\alpha\del_\mu\del_\nu h^{\mu\nu}-\square\del^\mu h_{\alpha\mu})^2/2$ as in \cite{blas}. Computing the two point amplitude in momentum space,  the gauge parameter disappears as expected and we have\footnote{The reader can check that $\lambda$-shifts $h_{\mu\nu}\rightarrow h_{\mu\nu}+\lambda \eta_{\mu\nu} h$ can be absorbed in redefinitions $a\to a(\lambda)$ and $b\to b(\lambda)$, see \eqref{a} and \eqref{b}, and 
$(T_{\mu\nu},J) \rightarrow (T_{\mu\nu}+\lambda T \eta_{\mu\nu},J \to J +\lambda T)$ which leaves $\tilde{T}_{\mu\nu}$ invariant but lead to $f_{D}(a,b)\rightarrow (1+\lambda D)^2f_{D}(a,b)$ and $\tilde{J}\rightarrow (1+\lambda D)\tilde{J}$, in such a way that \eqref{ampldiag} is invariant.},
\be 
\mathcal{A}(p)=-i T^{*\mu\nu}\mathcal{O}^{-1}_{\mu\nu,\alpha\beta}T^{\alpha\beta}=\frac{2i}{p^2}\left[\tilde{T}^{*\mu\nu}\tilde{T}_{\mu\nu}-\frac{|\tilde{T}|^2}{D-2}+\frac{D-2}{f_{D}(a,b)}|\tilde{J}|^2 \right] \label{ampldiag}.
\ee
The first two terms inside the brackets are precisely the same ones of the linearized Einstein-Hilbert (LEH) theory. Consequently, TDiff is consistent with any experimental results from LEH as far as the scalar contribution $\frac{D-2}{f_{D}}|\tilde{J}|^2$ is small enough. Notice that only positive deviations from LEH could be explained by a scalar contribution since $f_{D}>0$. In particular, if we take $\tilde{J}=c\, \tilde{T}$ we would have from \eqref{ampldiag}, in $D=4$, the following exceeding change in the deflection angle of the stars light by the sun: $\theta_{TDiff} = [1+4\, c^2/f_4(a,b)]\theta_{LEH}$.

Moreover, we can calculate the imaginary part of the residue of $\mathcal{A}(p)$ at $p^2\rightarrow 0$ and easily check that it is positive if $f_{D}(a,b)>0$. Since the source parametrization \eqref{source} is completely general, we have an explicitly covariant off-shell proof of unitarity of $\Lag (a,b)$.

\section{Massive scalar-tensor models in curved backgrounds}

This section is a preliminary step towards the construction of  possible generalizations of the massive scalar tensor theory discussed before, beyond flat backgrounds. Following the approach of \cite{buchbinder}, we investigate under which conditions we have the correct number of degrees of freedom in a curved background. See also \cite{hinter} and references therein and more recently \cite{bfh,hemily}.

\subsection{The constraints}

In $D=4$, our scalar-tensor models will contain a set of 11 independent variables, $h_{(\mu\nu)}$ and $\phi$. This number must be reduced to 6 since we need 5 for the massive spin-2 and 1 for the massive scalar. So we have to find out 5 constraints from the equations of motion.  We need to search for a vector and a scalar constraint. This is true also for any dimension $D\ge 3$.

We start writing the most general, up to second order in derivatives and quadratic in the fields, massive scalar-tensor model coupled to an external gravitational field via covariant derivatives and the addition of non-minimal coupling terms linear in curvatures. We begin with 14 arbitrary coefficients, including different mass terms for the tensor and scalar fields.
\begin{align}
\Lag=\Lag_{min}+ \dfrac{a_1}{2} R h_{\mu\nu}h^{\mu\nu}+\dfrac{a_2}{2} R h^2+\dfrac{a_3}{2} C^{\mu\alpha\nu\beta}h_{\mu\nu}h_{\alpha\beta}+\dfrac{a_4}{2} \tr^{\mu\beta}h_{\mu\nu}h^{\nu}{}_{\beta}+\dfrac{a_5}{2} \tr^{\mu\nu}h_{\mu\nu} h + \nn \\
+ \dfrac{b_1}{2} R\varphi^2 + \dfrac{b_2}{2} R h \varphi + \dfrac{b_3}{2} \tr^{\mu\nu}h_{\mu\nu}\varphi-\dfrac{m^2}{4}h^{\mu\nu}h_{\mu\nu}+c_1 \dfrac{m^2}{4}h^2+c_2 \dfrac{m^2}{2}h\varphi+c_3 \dfrac{m^2}{4}\varphi^2. \label{281}
\end{align}
where we define the traceless part of the Ricci curvature $\tilde{R}_{\mu\nu} = R_{\mu\nu}-\eta_{\mu\nu}\frac{R}D$ and the minimally coupled TDiff-like scalar tensor theory,
\bea
\Lag_{min}&=& \Lag_{a,b}^{\nabla} + \dfrac{x}{2}\varphi\nabla_\mu\nabla_\nu h^{\mu\nu} + \dfrac{y}{2}h\square \varphi + \dfrac{z}{4}\varphi\square\varphi,\label{min}\\
\Lag_{a,b}^{\nabla}&=&-\dfrac{1}{4}\nabla_\mu h^{\alpha\beta}\nabla^\mu h_{\alpha\beta}+\dfrac{1}{2}\nabla^\mu h^{\alpha\beta} \nabla_{\alpha}h_{\mu\beta}-\dfrac{a}{2}\;\nabla^\mu h \nabla^\nu h_{\mu\nu}+\dfrac{b}{4}\nabla_\mu h \nabla^\mu h.
\label{Ltdiff}
\eea
The equations of motion of \eqref{281} with respect to $h_{\mu\nu}$ are given by
\begin{align}
E_{\mu\nu}=\square h_{\mu\nu}-\nabla^\alpha(\nabla_\nu h_{\mu\alpha}+\nabla_\mu h_{\nu\alpha})+a \nabla_\mu\nabla_\nu h+a g_{\mu\nu} \nabla^\alpha\nabla^\beta h_{\alpha\beta}-b g_{\mu\nu}\square h + \nn \\+x\nabla^\mu\nabla^\nu \varphi + y \; g_{\mu\nu}\square \varphi + \mathbf{o}(h,\varphi)=0,
\end{align}
where $ \mathbf{o}(h,\varphi) $ denotes terms  with less than two derivatives. Explicitly, we have 
\begin{align}
\mathbf{o}&(h,\varphi)=  2 a_1 R h_{\mu\nu}+ 2 a_2 R g_{\mu\nu}h+2a_3 C_{\mu\alpha\nu\beta}h^{\alpha\beta}+a_4 (\tilde{R}_{\mu\beta}h_\nu{}^{\beta}+\tilde{R}_{\nu\beta}h_\mu{}^{\beta})\nn \\ & +a_5(\tilde{R}_{\mu\nu}h+g_{\mu\nu}R^{\alpha\beta}h_{\alpha\beta})  + b_2 g_{\mu\nu}R \varphi+b_3 \tilde{R}_{\mu\nu}\varphi-m^2 h_{\mu\nu}+c_1 \, m^2 g_{\mu\nu}h +c_2 \, m^2 g_{\mu\nu}\varphi . \label{ohphi}
\end{align}
The equation of motion of \eqref{281} with respect to  $\varphi$ is
\be
\Psi=x\;\nabla^\mu\nabla^\nu h_{\mu\nu}+y\;\square h+z\;\square \varphi + \mathbf{o}(h,\varphi)=0,
\ee
with
\be
\mathbf{o}(h,\varphi)= 2 b_1 R \varphi + b_2 R h+b_3 \tr_{\mu\nu}h^{\mu\nu}+c_3\;m^2\varphi+c_2\;m^2 h.
\ee
Similarly to what we have done in the flat space in subsection 3.1 we try to build a vector constraint via the linear combination:
\be 
\phi_\nu=\nabla^\mu E_{\mu\nu}+A_1\nabla_\nu g^{\alpha\beta}E_{\alpha\beta}+A_2\nabla_\nu \Psi=0.
\ee
One can show that
\bea
\phi_\nu &=&\nabla_\nu\square h\left[(a-b)+A_1(1+a-b D)+A_2 y\right] + \nabla_\nu \nabla^{\mu}\nabla^{\alpha}h_{\mu\alpha}\left[(a-1)+A_1(a D-2)+A_2 x\right]  \nn 
\\ &+& \nabla_\nu \square\varphi \left[(x+y)+A_1(x+y D)+A_2 z \right]+\cdots, \label{vec}
\eea
where the dots denote terms that contain less than two covariant derivatives. We can guarantee the existence of a vector constraint if we get rid of the second covariant derivatives\footnote{More precise and less restrictive is the requirement of getting rid of all second order {\bf time} derivatives only. For simplicity we will not investigate this possibility in the present work.}, namely,
\be
\begin{cases}(a-b)+A_1(1+a-b D)+A_2 y=0,\\(a-1)+A_1(a D-2)+A_2 x=0, \\(x+y)+A_1(x+y D)+A_2 z =0. \label{vectorsystem} \end{cases}
\ee
If $f_D\equiv f_{D}(a,b)\neq 0$, we obtain, 
\be
A_1=\dfrac{(b-a)x+(a-1)y}{x(1+a-bD)+y(2-aD)}, \qquad A_2=\dfrac{f_{D}}{x(1+a-bD)+y(2-aD)}, \label{AB}\ee
\be z=\dfrac{[b(D-1)-1]x^2+(D-2)y^2+2[(D-1)a-1]xy}{f_{D}}.\ee
The reader may worry about the denominators in \eqref{AB}, but it can be shown that if they vanish we necessarily have $f_{D}=0$. Notice that no non minimal terms are required for the existence of the vector constraint.

Now we look for the scalar constraint via the Ansatz:
\be 
\Omega=\nabla^\nu\phi_\nu + (A_3 m^2+A_4 R)g^{\mu\nu}E_{\mu\nu} + (A_5 m^2+ A_6 R)\Phi + A_7 \tr^{\mu\nu}E_{\mu\nu}=0.
\ee
The combination $ \nabla^\nu\phi_\nu $ is required in order to cancel the four-derivative terms. Collecting terms with two derivatives, 
\bea
\Omega &=& 2a_3 C_{\mu\alpha\nu\beta}\nabla^\mu\nabla^\nu h^{\alpha\beta}+\left[a_5+b_3 A_2+(2a_4+a_5 D)A_1+A_7\right]\!\tr_{\mu\nu}\square h^{\mu\nu} \nn\\ &+& \left(a+a_5+a A_7\right)\!\tr_{\mu\nu}\nabla^\mu\nabla^\nu h \nn \\ &+& \tr_{\mu\nu}\nabla^\mu\nabla^\nu \varphi \left[x(1+A_7)+b_3\right] + \tr_{\mu\nu}\nabla^\mu\nabla_\alpha h^{\alpha\nu} \left[2(a_4-1)-2 A_7\right] \nn \\ &+&
R\nabla^\mu\nabla^\nu h_{\mu\nu}\left[2a_1-\dfrac{2}{D}+A_4(a D-2)+A_6 x \right]\nn\\ &+& R\square\varphi\left[b_2+\dfrac{x}{D}+A_1b_2 D+2A_2 b_1+A_4(x+Dy)+A_6 z\right] \nn \\ &+& R\square h\left[2a_2+\dfrac{a}{D}+A_1(2a_2D+2a_1)+A_2 b_2 + A_4(1+a-bD)+A_6 y\right] \nn \\ &+&
m^2\nabla^\mu\nabla^\nu h_{\mu\nu}\left[-1+A_3(aD-2)+A_5x
\right]\nn\\ &+& m^2\square h\left[c_1+A_1(c_1 D-1)+A_2 c_2+A_3(1+a-bD)+A_5y\right] \nn \\
&+& m^2\square\varphi\left[c_2+A_1Dc_2+A_2c_3+A_3(x+Dy)+A_5z\right]+\dots = 0 \, .
\label{scalarc}
\eea
Once again dots stand for less than two derivatives. The first term of \eqref{scalarc} implies\footnote{Alternatively, we could have chosen to eliminate the Weyl tensor $C_{\mu\alpha\nu\beta}$, but this choice turns out to be more restrictive than the result we are going to get.} $a_3=0$. By using \eqref{AB} in the four terms with $\tr_{\mu\nu}=0$, i.e. second to fifth terms,  in \eqref{scalarc},  we can show that it is impossible to simultaneously get rid of these four terms. Consequently, we assume \textbf{Einstein spaces}:
\be 
\tr_{\mu\nu}=0 \label{einstein} .
\ee
The other six terms allow us to solve for the coefficients $A_3,A_4,A_5,A_6$, fixing the scalar constraint and determining two out of the seven coefficients $a_1,a_2,b_1,b_2,c_1,c_2$ and $c_3$. The conclusion is that regardless of the coefficients in the Lagrangian, see \eqref{281}, Einstein spaces are always required. This is also the case when $f_{D}=0$, not considered in \eqref{AB}, it contains the massive Fierz-Pauli \cite{buchbinder} and massive WTDiff \cite{bfh} theories. The later one corresponds to the FP case with the replacement $h_{\mu\nu} \to h_{\mu\nu} - (h/D + \phi ) g_{\mu\nu}$, see \cite{bfh}. It is  not difficult to see that equation \eqref{vectorsystem} is satisfied. Explicitly, for FP, $\varphi=0=A_1\Rightarrow \phi_\nu=\nabla^\mu E_{\mu\nu}=0$, whereas for WTDiff, $A_2=1/D$ and $A_1$ is redundant since $g^{\mu\nu}E_{\mu\nu}=0$ due to the Weyl symmetry, so that $\phi_\nu = \nabla^\mu E_{\mu\nu} - 1/D \nabla_\nu \Phi=0$. The scalar constraint requires Einstein spaces and non minimal couplings. If one adds extra non minimal terms including higher powers in the curvature and negative powers of the mass, like e.g. $R^4/m^2$, it turns out \cite{buchbinder} that one can lift the requirement of Einstein spaces. Moreover, it is remarkable \cite{bernard,b2} that the non minimal coefficients  can be obtained from the ghost free massive gravity theory of \cite{drgt,hs2}  by eliminating the fiducial metric $f_{\mu\nu}$ order by order in $1/m^2$ in terms of the background metric $g_{\mu\nu}^{(0)}$ around which one expands the dynamic metric $g_{\mu\nu}=g_{\mu\nu}^{(0)}+h_{\mu\nu}$ or by the expansion, see \cite{b3}, of the bimetric model of \cite{hs3} in the massive gravity limit. Here we only work with non singular powers of the mass and stick to Einstein spaces.

Back to our scalar-tensor models ($f_{D}\neq 0$), that describe 6 degrees of freedom in $D=3+1$,  let us find out an explicit expression for the curved version of the massive extension \eqref{ltdm}. Fixing $x=a-1$, $y=a-b$, $z=2a-b-1$, $c_1=b$, $c_2=-y$,  $c_3=-z$, and solving  \eqref{vectorsystem} and \eqref{scalarc} for Einstein spaces, we obtain a consistent family of curved background generalizations of \eqref{ltdm},
\bea
\Lag_m &=& \Lag^{\nabla}_{TDiff}  -\dfrac{m^2}{4}(h_{\mu\nu}h^{\mu\nu}-b\;h^2)+\dfrac{a-1}{2}\varphi\nabla_\mu\nabla_\nu h^{\mu\nu} + \dfrac{a-b}{2}h(\square-m^2)\varphi + \nn \\ &+&\dfrac{2a-b-1}{4}\varphi(\square-m^2)\varphi + \dfrac{a_1}{2} R h_{\mu\nu}h^{\mu\nu} + \dfrac{a_2}{2}R h^2 + \dfrac{b_1}{2}R \varphi^2 + \dfrac{a_1+a_2+b_1-1/(2D)}{2}R h \varphi \nn\\ \label{lmtdiffeinstein}
\eea
where $\Lag^{\nabla}_{TDiff}$ is given in \eqref{Ltdiff}. Although $a_1,a_2$ and $b_1$ are free, there is no way of avoiding non minimal couplings just like in the Diff and WTDiff cases.

\subsection{Massless symmetries}

It is expected that local symmetries of the massless theory play a key role in the derivation of constraints in massive theory as in the usual flat space Maxwell-Proca and Fierz-Pauli models. So now we investigate the interesting interplay between the massless symmetries and the constraints.

From \eqref{281} with $m=0$, we require its invariance under linearized generalized Diff
%\footnote{This is a generalized Diff transformation because it depends on $D$ parameters, being Diff a sub-case when $c=0=k$.} 
\be 
\delta h_{\mu\nu}= \nabla_\mu \xi_\nu+\nabla_\nu \xi_\mu + k_1\; g_{\mu\nu} \nabla\cdot\xi,  \qquad \delta  \varphi=k_2\;\nabla\cdot\xi, \qquad (k_1,k_2)\in\mathbb{R}. \label{gendiff}
\ee
where $k_1,k_2$ must be determined. By explicit computation, we reach the same conclusions as before: \textbf{Einstein spaces are required}, and $a_3=0$. Moreover, we still have: 
\begin{align}
&\delta\Lag=\dfrac{R}{D}h_{\mu\nu}\nabla^\mu\xi^\nu[2 (a_1 D - 1) ]+\dfrac{R}{D}h\nabla\cdot\xi\left[a+a_1 k_1 D+a_2(k_1 D+2)D+\dfrac{b_2}{2}k_2 D\right]\nn \\
& + \dfrac{R}{D}\varphi\nabla\cdot\xi\left[x+b_1 k_2 D+\dfrac{b_2}{2}(k_1 D+2)D\right] + h \square \nabla\cdot\xi \left[ \dfrac{a}{2}(k_1+2) - \dfrac{b}{2}(2+k_1 D)+\dfrac{k_1}{2}+\dfrac{y}{2}k_2\right]  \nn \\
&+\nabla_\mu \nabla\cdot\xi \nabla_\nu h^{\mu\nu}\left[1+k_1-\dfrac{a}{2}(k_1 D+2)-\dfrac{x}{2}k_2\right] + \varphi \square \nabla\cdot\xi\left[\dfrac{x}{2}(2+k_1)+\dfrac{y}{2}(2+k_1 D)+\dfrac{z}{2}k_2\right]. \label{system}
\end{align}

So we do have TDiff symmetry ($\nabla\cdot\xi=0$)
 in the massless version of \eqref{281} on Einstein spaces if
 $(a_1,a_3)=(1/D,0)$. Notice however, that the vanishing of the last three brackets supplies us with the same three equations \eqref{vectorsystem} necessary for the vector constraint  by identifying $(A_1,A_2)=(k_1/2,k_2/2)$. Thus, Diff symmetry at $m=0$ implies the vector constraint. Conversely and more interesting, the existence of the vector constraint implies Diff symmetry at $m=0$ in the flat limit. In particular, there is no hope of a consistent massive scalar tensor model whose massless limit is only TDiff invariant in the flat space. Although TDiff is the minimal symmetry for massless spin-2 particles, the massive case seems to require full Diff in the massless sector. Finally, we stress that, as in the $f_D=0$ case \cite{buchbinder}, within the truncations made here we can have a massive scalar tensor model in Einstein spaces with a scalar and a vector constraint without any local symmetry at $m=0$. The requirement of those symmetries imposes further constraints on the non minimal coefficients. For instance, in order that we have Diff symmetry in the massless limit of \eqref{lmtdiffeinstein} we must have,
\be a_1= \frac 1D \quad ; \quad  b_1=a_2+\frac aD-\frac 1{2D} \label{a1b1} \ee
We recall the reader 
 that in Einstein spaces the
scalar curvature $R$  is constant and plays a similar role  of $m^2$. There are special values of $m^2/R$ for which vector and scalar symmetries show up which are under investigation.  

At last, we point out that the existence of constraints is equivalent to a simple counting of degrees of freedom, a necessary but not a sufficient condition for the propagation of physical particles in a curved background,  unitarity and causality should be further investigated even within the truncations made here.

\section{Conclusion}

We have made the Kaluza-Klein dimensional reduction from the $D+1$ TDiff model \eqref{LTD} to $D$ dimensions and have obtained, for arbitrary values of $(a,b)$, the simple expression \eqref{ldtd} in terms of gauge invariant fields. The connection between the original tensor field $h_{\mu\nu}$ and $\HTD$ is nontrivial in general but we have shown the existence of unitary gauges in all cases. The case $f_{D+1}(a,b)=0$ contains the Diff model, linearized general relativity, and WTDiff,  linearized unimodular gravity. Both cases lead to massive spin-2 particles. All remaining cases correspond to a massive scalar tensor model. We point out that there is no guarantee, in general, that one will be able to write down the reduced model in terms of field combinations involving the Stückelberg fields invariant under all local symmetries. It may happen, see \cite{bdg}, that there is no combination of fields large enough to accommodate all symmetries, so part of them must be realized dynamically by combining the different terms in the Lagrangian.

In subsection 3.1.2 we have investigated the nontrivial issue of gauging away Stückelberg fields at action level and showed that the completeness (uniqueness) criterion of \cite{moto} fits well with the unitary requirements for a massive spin-2 theory. The message is that we have to be careful and avoid gauge conditions with residual symmetries.

In subsection 3.2  we have checked, in all cases, that the massless limit of the reduced model \eqref{ldtd} contains the same number of degrees of freedom as the corresponding massive theory, thus providing a smooth massless limit.

In section 4 we obtain the massive scalar tensor model \eqref{ltdm}  via gauge fixing of the reduced model. The scalar and tensor degrees of freedom can not be decoupled by any local field redefinition. By means of a helicity decomposition we have identified the different helicity modes. We have also worked out the equations of motion and the necessary vector and scalar constraints for a massive scalar tensor theory. The coupling to arbitrary sources have allowed us to prove unitarity in a covariant way and to show that the scalar field  can only worsen the mass discontinuity problem \cite{vdvz}.

The massless version of \eqref{ltdm}, i.e., \eqref{lm0} is invariant under Diff and not TDiff, similarly the massless limit of the WTDiff reduced model investigated in \cite{bfh} is invariant under WDIFF and not WTDiff. In section 5 we show, starting from a more general Ansatz, that in order to have the vector and the scalar constraints in a second order scalar tensor model we do need Diff symmetry in the massless sector in flat space. So the absence of a consistent Lorentz covariant mass term for spin-2 particles in the pure tensor TDiff model \eqref{L2} seems to be more general. Apparently, although TDiff is the minimal symmetry  \cite{van} for massless spin-2 particles, the massless sector must be Diff invariant in order that the mass terms produce the necessary number of constraints for the elimination of non physical degrees of freedom.

We have found a non local correspondence \eqref{redeftdeh} between the pure tensor model \eqref{L2} and \eqref{3.15} leading to a parametrization \eqref{source} of the non conserved source allowing an explicitly covariant proof of unitarity of \eqref{L2} and a clear separation of the spin-2 and spin-0 contributions to the two point amplitude of the pure tensor TDiff model, see \eqref{ampldiag}.
The relationship \eqref{redeftdeh} takes us also from the WTDiff model to the Einstein-Hilbert theory. We wonder whether there would be any nonlinear generalization connecting unimodular gravity to general relativity.

Finally, in section 5 we investigate linearized massive scalar tensor theories in curved backgrounds. We go beyond the scalar tensor theory obtained via dimensional reduction and start from a rather general second order, in derivatives, Ansatz with 14 parameters and non minimal terms linear in the curvature. The requirement of a vector plus a scalar constraint leads to the equations \eqref{vecto} and \eqref{scalarc}. Instead of a detailed analysis of a complicate system of equations (under investigation) we show that the specific flat space massive model \eqref{ltdm} does admit a family of curved background extensions on Einstein spaces  with three free non minimal parameters. If we demand Diff symmetry at $m=0$, only one parameter, see \eqref{a1b1}, remains arbitrary. The natural way of lifting the Einstein spaces demand is to include  non minimal terms with higher powers in the curvature and negative powers in $m^2$ as in \cite{buchbinder}. We may go even ahead that perturbative approach and try to formulate a massive scalar tensor gravity along the lines of \cite{drgt} or even a scalar bi-tensor theory generalizing \cite{hs3}. This is beyond the scope of the present work.

\section{Acknowledgements}

The work of D.D. is partially supported by CNPq  (grant 306380/2017-0). R. R.  is supported by a research grant (29405) from VILLUM FONDEN, and grateful for the support by FAPESP (grant 2018/24767-2) and CNPq. D.D. is grateful to Gabriel B. de Gracia for earlier (2016) discussions on the dimensional reduction of TDiff models. We thank the anonymous referee for suggestions.

\section{Appendix: Projection operators} 

We can define \textbf{projection operators} acting on vector fields according to their longitudinal and transverse components, 
\be
\omega_{\mu\nu}=\frac{\del_{\mu}\del_{\nu}}{\square} \quad ; \quad
\theta_{\mu\nu}=\eta_{\mu\nu}-\omega_{\mu\nu}.
\label{omegatheta}
\ee
They satisfy the algebra and the closure relation below, 
\be \theta_{\mu\alpha}\theta^{\alpha\nu}=\theta_{\mu}^{\nu} \qquad \quad ; \quad \omega_{\mu\alpha}\omega^{\alpha\nu}=\omega_{\mu}^{\nu} \quad ; \quad \omega_{\mu\nu}\theta^{\nu\alpha}=0 \quad ; \quad \delta^{\mu}_{\nu}=\theta^{\mu}_{\nu}+\omega^{\mu}_{\nu}, \label{completeza}\ee

From the vector projection operators we can construct a basis for operators acting on rank-2 tensors. Here we use the projection operators of \cite{pvn73}. Notice that a slightly different basis was employed before by \cite{rivers}. Let us limit ourselves to rank-2 symmetric tensors:

\be
(P_{ss}^{(2)})^{\mu\nu\alpha\beta} \equiv \frac{1}{2}\left(\theta^{\mu\alpha}\theta^{\nu\beta} + \theta^{\mu\beta}\theta^{\nu\alpha}\right)-\frac{1}{D-1}\theta^{\mu\nu}\theta^{\alpha\beta} , \label{Pss2}
\ee
\be 
(P_{ss}^{(1)})^{\mu\nu\alpha\beta} \equiv \frac{1}{2}\left(\theta^{\mu\alpha}\omega^{\nu\beta} + \theta^{\mu\beta}\omega^{\nu\alpha} + \theta^{\nu\alpha}\omega^{\mu\beta} + \theta^{\nu\beta}\omega^{\mu\alpha}\right), \label{Pss1} 
\ee
%\be
%(P_{aa}^{(1)})^{\mu\nu\alpha\beta} \equiv \frac{1}{2}\left(\theta^{\mu\alpha}\omega^{\nu\beta} - \theta^{\mu\beta}\omega^{\nu\alpha} - \theta^{\nu\alpha}\omega^{\mu\beta} + \theta^{\nu\beta}\omega^{\mu\alpha}\right) \label{Paa1}
%\ee
%\be
%(P_{as}^{(1)})^{\mu\nu\alpha\beta} \equiv \frac{1}{2}\left(\theta^{\mu\alpha}\omega^{\nu\beta} + \theta^{\mu\beta}\omega^{\nu\alpha} - \theta^{\nu\alpha}\omega^{\mu\beta} - \theta^{\nu\beta}\omega^{\mu\alpha}\right) \label{Pas1}
%\ee
%\be
%(P_{sa}^{(1)})^{\mu\nu\alpha\beta} \equiv \frac{1}{2}\left(\theta^{\mu\alpha}\omega^{\nu\beta} - \theta^{\mu\beta}\omega^{\nu\alpha} + \theta^{\nu\alpha}\omega^{\mu\beta} - \theta^{\nu\beta}\omega^{\mu\alpha}\right) \label{Psa1}
%\ee
\be 
(P_{ss}^{(0)})^{\mu\nu\alpha\beta} \equiv \frac{\theta^{\mu\nu}\theta^{\alpha\beta}}{D-1}
\qquad , \qquad
(P_{ww}^{(0)})^{\mu\nu\alpha\beta} \equiv \omega^{\mu\nu}\omega^{\alpha\beta}. \label{Pww0}
\ee
%\be 
%(P_{aa}^{(0)})^{\mu\nu\alpha\beta} \equiv \frac{1}{2}\left(\theta^{\mu\alpha}\theta^{\nu\beta}-\theta^{\mu\beta}\theta^{\nu\alpha}\right) \label{Paa0}
%\ee
In this case, it is necessary to include transition operators
\be
(P_{sw}^{(0)})^{\mu\nu\alpha\beta} \equiv \frac{\theta^{\mu\nu}\omega^{\alpha\beta}}{\sqrt{D-1}} 
\qquad,\qquad
(P_{ws}^{(0)})^{\mu\nu\alpha\beta} \equiv \frac{\omega^{\mu\nu}\theta^{\alpha\beta}}{\sqrt{D-1}}. \label{Psw0}
\ee
Both projection and transition operators satisfy the following algebra
\be 
 P_{ij}^{(r)}P_{kl}^{(q)} = \delta^{rq}\delta_{jk} P_{il}^{(r)}, \label{algebra}
\ee
and the closure relation is given by 
\be 
1_S^{\mu\nu\alpha\beta}\equiv  \frac{1}{2}\left(\eta^{\mu\alpha}\eta^{\nu\beta}+\eta^{\nu\alpha}\eta^{\mu\beta}
\right)=\left(P_{ss}^{(2)} + P_{ss}^{(1)} + P_{ss}^{(0)} + P_{ww}^{(0)}\right)^{\mu\nu\alpha\beta} , \label{identidade simetrica}
\ee
%\be
%A^{\mu\nu\alpha\beta}=\left(P_{aa}^{(1)} + P_{aa}^{(0)}\right)^{\mu\nu\alpha\beta}  = \frac{1}{2}\left(\eta^{\mu\alpha}\eta^{\nu\beta}-\eta^{\nu\alpha}\eta^{\mu\beta}\right)	\label{identidade antissimetrica}
%\ee
The trace of projection operators $P_{ij}^{(s)}$ is related to spin $s$ through: ${\rm trace}=2s+1$.

\end{document}